\documentclass[journal]{IEEEtran}

\IEEEoverridecommandlockouts
\usepackage{bm}
\usepackage{cite}
\usepackage{url}
\usepackage{float}
\usepackage{amsmath}
\usepackage{color}
\usepackage{amssymb,amsthm,graphicx,subfigure,cite,color,enumerate,algorithm,algorithmic}
\usepackage{diagbox}
\usepackage[table]{xcolor}
\usepackage{booktabs}
\usepackage{stfloats}
\usepackage{CJK}
\usepackage{subfigure}
\usepackage{balance}
\usepackage{footnote}
\usepackage{makecell}
\date{}
\makesavenoteenv{align}


\allowdisplaybreaks[4]

\begin{document}

\title{Vision-Aided Blockage Avoidance in UAV-assisted V2X Communications}

\author{Ruijin~Ding,~\emph{Student Member, IEEE}, Weihua~Xu,~\emph{Student Member, IEEE}, Wanmai~Yuan, and 
        Feifei~Gao,~\IEEEmembership{Fellow,~IEEE}
\thanks{R.~Ding, W.~Xu and F.~Gao are with Institute for Artificial Intelligence Tsinghua University (THUAI),
State Key Lab of Intelligent Technologies and Systems, Beijing National Research Center for Information Science and Technology (BNRist),
Department of Automation, Tsinghua University, Beijing 100084, P.R. China. (email: drjdtc@outlook.com, xwh19@mails.tsinghua.edu.cn, feifeigao@ieee.org).}
\thanks{W. Yuan is with Information Science Academy of CETC. (email: yuanwanmai7@163.com)}
}

\maketitle

\begin{abstract}
The blockage is a key challenge for millimeter wave communication systems, since these systems mainly work on line-of-sight (LOS) links, and the blockage can degrade the system performance significantly.
It is recently found that visual information, easily obtained by cameras, can be utilized to extract the location and size information of the environmental objects, which can help to infer the communication parameters, such as blockage status. In this paper, we propose a novel vision-aided handover framework for UAV-assisted V2X system, which leverages the images taken by cameras at the mobile station (MS) to choose the direct link or UAV-assisted link to avoid blockage caused by the vehicles on the road. 
We propose a deep reinforcement learning algorithm to optimize the handover and UAV trajectory policy in order to improve the long-term throughput. Simulations results demonstrate the effectiveness of using visual information to deal with the blockage issues. 
\end{abstract}


%
\IEEEpeerreviewmaketitle

\section{Introduction}
In future wireless networks, high-frequency signals like 5G millimeter wave (mmWave) and 6G sub-terahertz, are essential for the communication service with large bandwidth. However, high-frequency communications mainly rely on line-of-sight (LOS) links, which are very sensitive to the blockage. When the LOS link between the user and the base station (BS) is blocked, the received signal power will suffer from severe attenuation, leading to severe reduction in the signal-to-noise-ratio (SNR). The attenuation of the signal can deteriorate the quality of experience (QoS), and the reconstruction of the LOS link is reactive. As a result, it is important for the high-frequency communication to maintain the LOS link proactively, which requires the capability of sensing the surrounding environment~\cite{andrews2016modeling}.

Recently, integrated sensing and communication (ISAC) has drawn increasing attention, since it can utilize the sensors, such as cameras, Radars, and LIDARs, equipped at the intelligent terminals to assist communication. The sensory data is the valid out-of-band information that can indicate the spatial characteristics of the surrounding environment.
In \cite{xu20203d}, the environmental information is obtained from 3D scene reconstruction based on images, and is used to assist beamforming.
In \cite{xu2021deep}, Xu \emph{et al.} utilize the environmental 3D images to predict the channel covariance matrix.
In \cite{charan2021vision}, Charan \emph{et al.} propose a bimodal machine learning solution to predict future link blockage based on consecutive RGB images taken at the BS and mmWave beams, which enables proactive handover between BSs. However, the handover between terrestrial BSs may still face blockage, for example, on a congested road. 

In this paper, we propose a vision-aided handover framework for UAV-assisted V2X communications, where the mobile station (MS) leverages the images taken by the onboard cameras to choose the direct link or the UAV-assisted link to connect the BS. Meanwhile, the UAV design its trajectory to obtain better channel for the MS. The images are taken at the MS because it is more convenient than at the BS with the gradual popularization of smart phones and automatic driving. Besides, taking images at the BS may violate the privacy of customers. The main contributions is summarized as follows:
\begin{itemize}
  \item We utilize the 3D object detection technique to extract the size and the location information from the images taken by the cameras equipped at the MS, and embed the extracted information into the original images by converting the RGB channel to length, width, and height channel.
  \item We use 2D Gaussian distribution to describe the size and the location information of all vehicles with the expected value denoting the 2D coordinates and the standard deviation denoting the size information.
  \item We develop a deep reinforcement learning (DRL) algorithm, VQMIX, to enable the MS to choose the direct link or UAV-assisted link and to enable the UAV to design its trajectory, for improving the communication performance.
\end{itemize}

%
%
%
%



\section{System Model and Problem Formulation}
\subsection{System Model}
\begin{figure}
  \centering
  \includegraphics[width=0.45\textwidth]{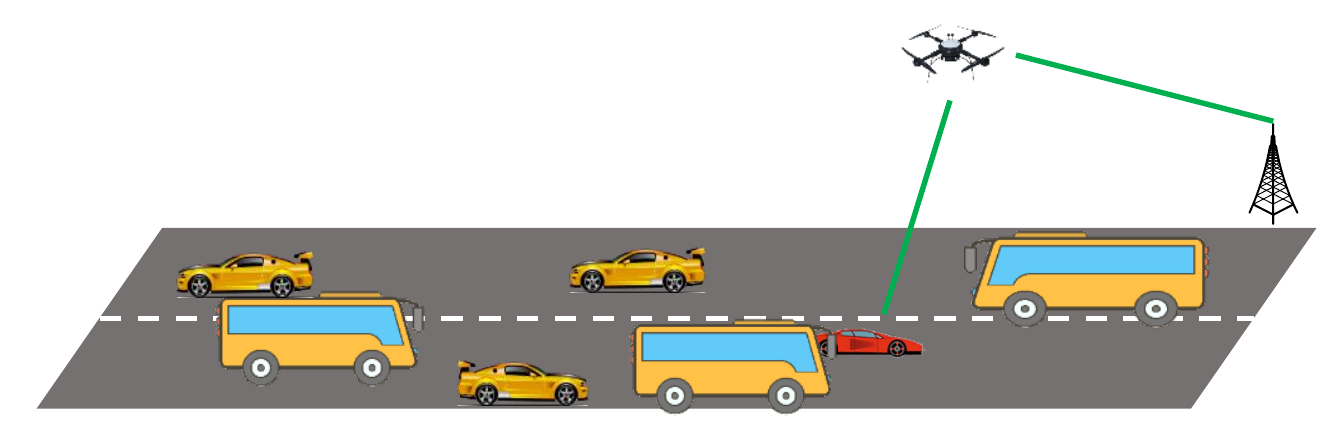}
  \caption{The V2X communication scenario.}
  \label{fig:system}
\end{figure}

As shown in Fig.~\ref{fig:system}, we consider a downlink orthogonal frequency division multiplexing (OFDM) V2X communication system with a single BS and a mobile station (MS). The MS is the target vehicle that moves along the right side of the road. The location of the MS is denoted by $\bm{w}^{M}$. There are many other vehicles, such as cars, vans and buses, that can be possible blockage objects. The vehicles follow the right-side traffic rule along the road. The BS is located at the roadside with the antenna height $h^{B}$. The communication link between the BS and MS can be blocked by the other vehicles on the road. There is a UAV acting as a data relay between the BS and the MS when the blockage happens. The flying altitude of the UAV is $h^{U}$, and the horizontal location of the UAV is denoted by $\bm{w}^{U}\in\mathbb{R}^{2\times 1}$. The UAV has a constant flying speed $V$. In order to obtain better communication service, the MS can choose the direct link to the BS or the UAV-assisted link, while the UAV can design its trajectory to guarantee the LoS links between the UAV and the MS as well as the LoS links between the UAV and the BS. 

We assume that the driving time of the MS on this road is $T$ time slots, and the duration of each time slot is $\delta_t$. At each time slot, the MS should choose UAV or the BS to access, and the UAV should choose the flying direction. The duration of each time slot $\delta_t$ is short enough such that the flying direction of the UAV keeps unchanged within a time slot. 

Since the objects on the road and around can reflect, scatter and diffract the signal, there are many paths between the transmitter and the receiver. 
We adopt the geometric channel model \cite{channel} to express the downlink channel between the BS and the MS for the $k$-th subcarrier at time slot $t$
\begin{equation}\label{eq:channel}
  H^{BM}_{k}(t) = \sum_{n=0}^{N-1}\sum_{l=1}^{L}\alpha_{l}^{BM}(t)e^{-j\frac{2\pi k}{K}n}d(nT_{s}-\tau^{BM}_{l}(t)),
\end{equation}
where $N$ is the length of cyclic prefix in the OFDM system, $L$ is the number of paths, $\alpha_{l}^{BM}(t)$ is the complex gain of the $l$-th path between the BS and the MS at time slot $t$, $K$ is the number of subcarriers, $d(\cdot)$ represents the pulse shaping filter, $T_{s}$ denote the sampling interval, and $\tau^{BM}_{l}(t)$ is the time delay of the $l$-th path. Similarly, the channels between the BS and the UAV as well as those between the UAV and the MS at the $k$-th subcarrier can also be calculated by \eqref{eq:channel}, and are denoted by $H^{BU}_{k}(t)$ and $H^{UM}_{k}(t)$, respectively. 
The downlink capacity between the BS and the MS can be calculated as 
\begin{equation}\label{eq:rbm}
  R^{BM}(t) = B_{k} \sum_{k=1}^{K}\log_{2}\left(1+\frac{P^{B}_{k}|H^{BM}_{k}(t)|^2}{n_{0}B_{k}}\right),
\end{equation}
where $B_{k}$ is the bandwidth of each subcarrier, $P^{B}_{k}$ is the transmit power of the BS on the $k$-th subcarrier, and $n_0$ is the noise power spectral density.
Similarly, the downlink capacity between the BS and the UAV and between the UAV and the MS can be calculated by \eqref{eq:rbm}, and denoted by $R^{BU}(t)$ and $R^{UM}(t)$.
When the MS choose the UAV-assisted link, the capacity of this link are
\begin{equation}
  R^{BUM}(t) = \min\{R^{BU}(t), R^{UM}(t)\}.
\end{equation} 
As a result, the capacity of the considered system is
\begin{equation}
  R^M(t) = a^{M}(t)R^{BUM}(t)+(1-a^{M}(t))R^{BM}(t),
\end{equation}
where $a^M(t)$ is a binary indicator that represents whether the MS chooses the direct link or the UAV-assisted link at time slot $t$, i.e., $a^M(t)=0$ means the MS chooses the direct link and $a^M(t)=1$, otherwise.

\subsection{Problem Formulation}

The objective of the considered system is to maximize the long-term throughput by making sequential decisions, i.e., the MS chooses the direct link or UAV-assisted link and the UAV designs its trajectory at the same time. Note that the UAV-assisted link can lead to communication delays, and the problem can be mathematically formulated as 
\begin{align}\label{pro:dis}
  \max\limits_{a^M(t),\bm{a^{U}}(t)} \quad &\sum_{t=0}^{T}R^{M}(t)-a^M(t)C\\
  \mbox{s.t.}\quad \ \ \quad & \bm{w}^{U}(t+1)-\bm{w}^{U}(t)=\bm{a^{U}}(t)V\delta_{t},\tag{\ref{pro:dis}{a}}\label{eq:move}\\
  & \bm{w}^{U}(0)=\bm{w}^{U}_0,\tag{\ref{pro:dis}{b}}\label{eq:start}\\
  & \bm{b}_{l}\leq\bm{w}^{U}(t)\leq \bm{b}_u,\tag{\ref{pro:dis}{c}}\label{eq:road}
\end{align}
where $C$ denotes the UAV-assisted link cost, $\bm{a^{U}}(t)$ denotes the flying direction of the UAV at time slot $t$, \eqref{eq:move} is the movement function, $\bm{w}^{U}_0$ denotes the initial location of the UAV, $\bm{b}_{l}$ and $\bm{b}_u$ are the lower and upper boundary coordinates of the considered road.
\section{Computer Vision Aided Vehicle Detection}\label{sec:cv}
\begin{figure}
  \centering
  \includegraphics[width=0.49\textwidth]{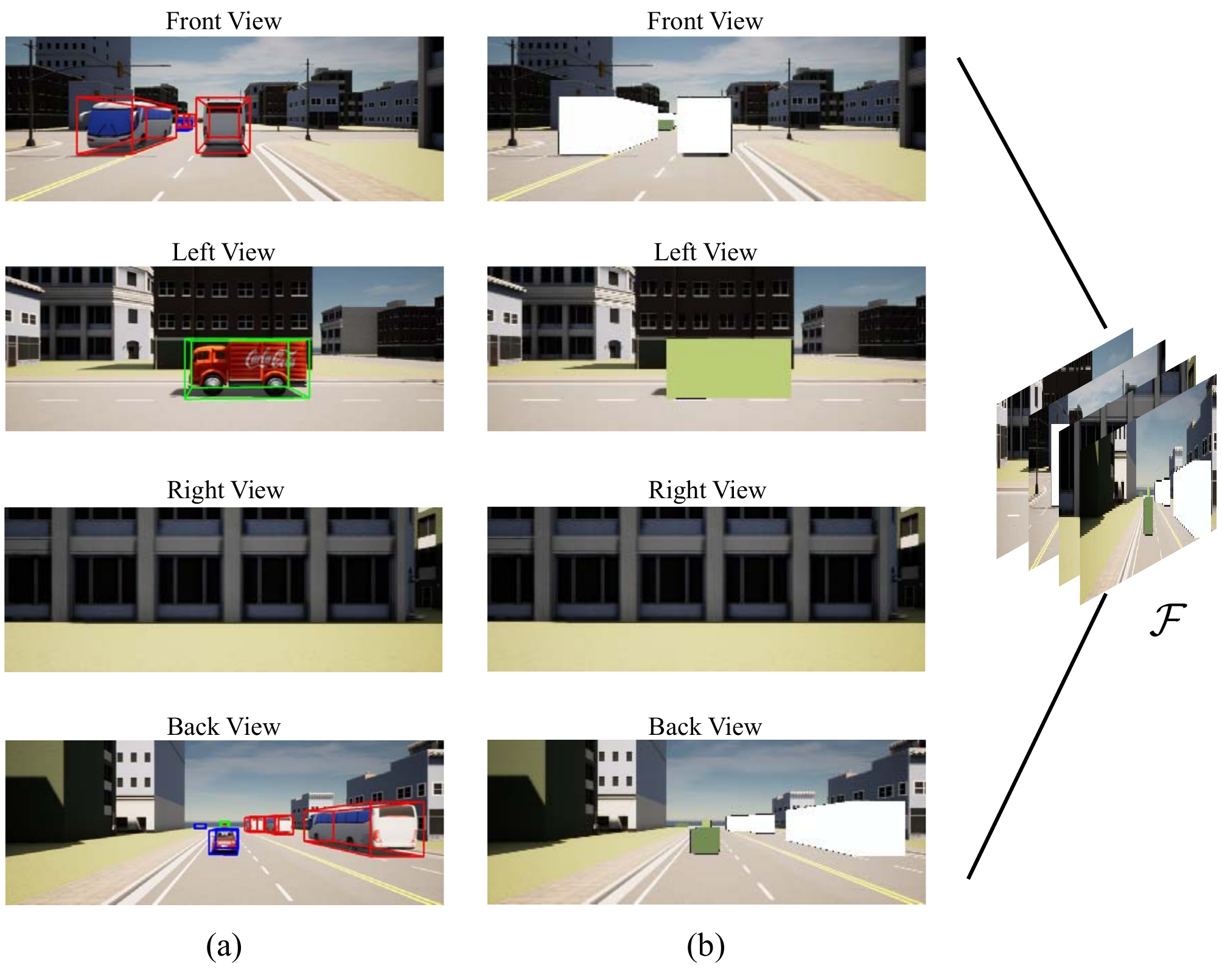}
  \caption{The detected vehicles are replaced by the size information.}
  \label{fig:projection}
\end{figure}

The channel is determined by the environmental objects, such as buildings, roads, and vehicles. To capture the feature of the environment, the MS is equipped with four monocular cameras to take the environmental images from front, back, left and right views, respectively. At each time slot, the MS takes four images through the cameras and make decisions based on the images.
Compared to the static objects, i.e., the buildings and the road, the dynamic objects, i.e., the vehicles, are the main factor leading to the dynamic changes of the channel. As a result, we should focus on the vehicles in the images. 
With the monocular 3D object detection \cite{3dod}, the sizes, locations and orientations of the vehicles can be extracted from the images. As shown in Fig.~(\ref{fig:projection}a), the MS adopts a single-stage monocular 3D object detection technique, i.e., SMOKE \cite{smoke}, to obtain the 3D bounding box of the vehicles in each image.

However, the pixel area of the vehicles in the original images may be inconsistent with the actual size of the vehicles, which is an inherent shortcoming of the 2D images. For example, a small vehicle may still look very large, if it is very close to the MS. 
Besides, whether a vehicle will block the direct link between the BS and the MS mainly depends on the location, size and orientation information of this vehicle. In contrast, the color information of this vehicle is unimportant and redundant, since the vehicle color has little impact on the signal propagation. 
Therefore, we convert the vehicle color information to the vehicle size information. We treat each vehicle as a cuboid, the size information consists of the length, width and height, which can be obtained by the 3D object detection technique, SMOKE. Note that the color information consists of three RGB channels, we can replace the RGB channel data of the pixels in the bounding box of each detected vehicle with the length, width and height of the vehicle. The length, width and height of the detected vehicle $k$ are denoted by $l^{V}_{k}$, $w^{V}_{k}$, $h^{V}_{k}$, respectively. As shown in Fig.~\ref{fig:projection}, we utilize the tuple $(-255\frac{l^{V}_{k}}{L_{\max}},-255\frac{w^{V}_{k}}{W_{\max}},-255\frac{h^{V}_{k}}{H_{\max}})$ to replace the corresponding RGB channel data of the detected vehicle pixels, where $L_{\max}$, $W_{\max}$ and $H_{\max}$ are the maximum length, width and height of all possible vehicles respectively. The minus sign is used to distinguish the vehicles from the background. As a result, we get four modified ``images'' with the size information of the detected vehicles.
At each time slot, we stack the four modified ``images'' together to formulate a feature map $\bm{\mathcal{F}}(t)$ with $4\times 3=12$ channels.

\section{Algorithm}

\begin{figure*}
  \centering
  \includegraphics[width=0.8\textwidth]{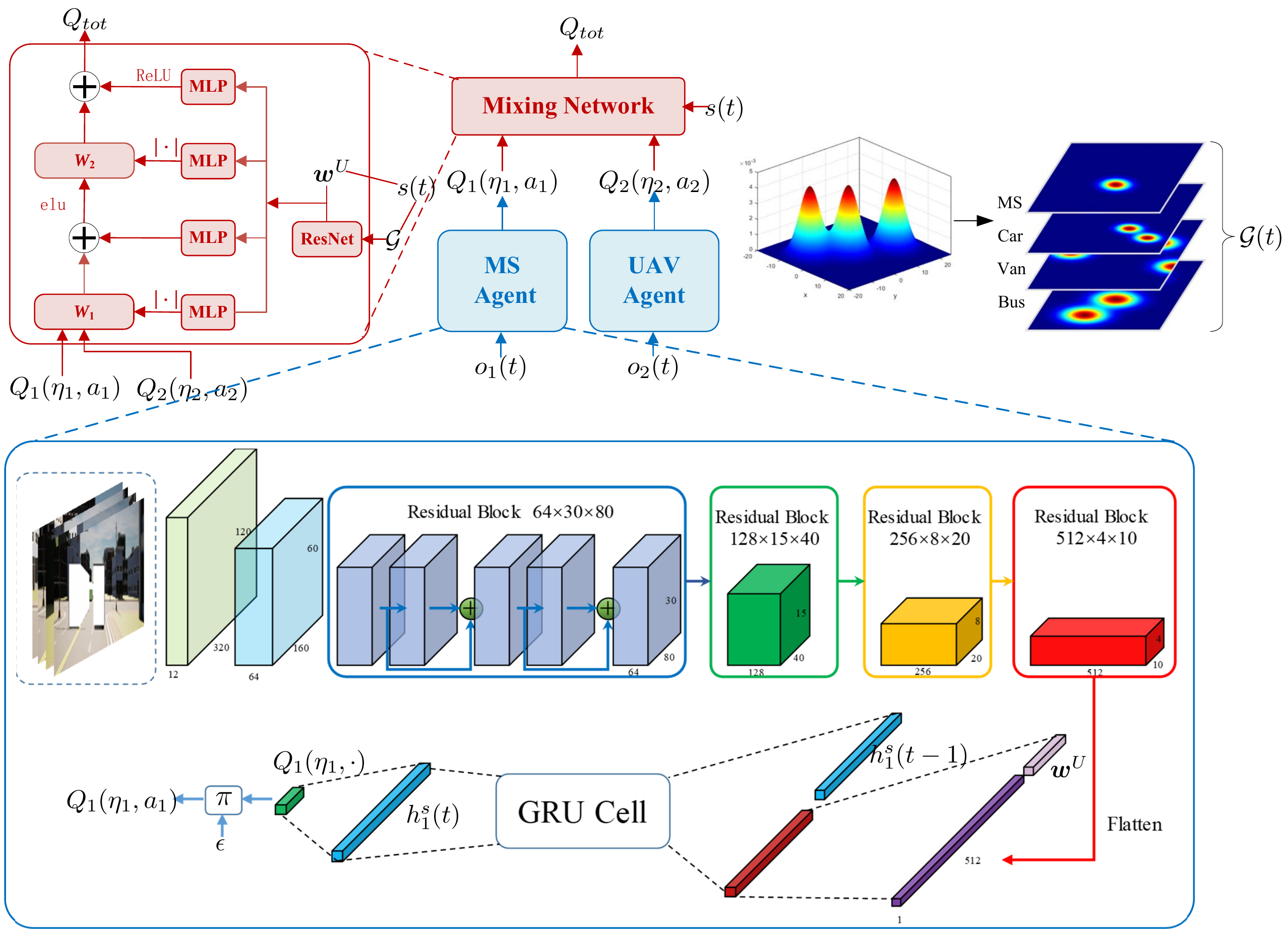}
  \caption{Illustration of the VQMIX.} 
  \label{fig:vqmix}
\end{figure*}

We choose the cooperative multi-agent reinforcement learning algorithm, QMIX \cite{qmix}, as the start point of the proposed algorithm, VQMIX. As shown in Fig.~\ref{fig:vqmix}, VQMIX aim to learn a joint action-value function $Q_{tot}(\bm{\eta},\bm{a})$, where $\bm{\eta}$ is the joint observation-action history and $\bm{a}$ is the joint action of all agents. Each agent has an individual action-value function $Q_i(\eta_i,a_i)$, where $\eta_i$ and $a_i$ are the observation-action history and action of agent $i$, respectively. Each agent has a GRU cell to learn from the observation-action history. 
There is a mixing network to take the individual action-value function of each agent $Q_i$ as the input and output the $Q_{tot}$. 
The weights of the mixing network are generated by some separate hypernetworks and activated by absolute activation function. Each hypernetwork takes the state $s(t)$ as the input, and outputs the weights of one layer of the mixing network.

The proposed VQMIX is for the downlink OFDM communication system, where the MS and the UAV are regarded as two agents. The MS chooses the communication links and the UAV design its trajectory to maximize the long-term throughput.
\subsection{Observation Space}
Each agent can only make decisions based on its available information, i.e., the observation. 
\subsubsection{MS}
As mentioned in Section~\ref{sec:cv}, the MS is equipped with four cameras to take photos of the surrounding environment. The MS utilizes 3D object detection technique to detect the on-road vehicles, and replace the color information of the detected vehicles with the size information of them to formulate the feature map $mathcal{F}$.

Besides, the location information are important for the MS to make decision. As a result, the observation of the MS is 
\begin{equation}
  o^M(t) = \{\mathcal{F}(t),\bm{w}^M(t)\}.
\end{equation}

\subsubsection{UAV}


The locations of vehicles are available for the UAV, since the UAV flies in the air and can observe the vehicles on the road. However, the number of vehicles on the road is constantly changing, i.e., the length of the location information is variable, which is not trackable for the neural network. We embed the location information of vehicles into a 2D plane. Furthermore, the size information is also important. In order to combine the location and size information, 2D Gaussian distribution is adopted. For vehicle $k$, we have
\begin{equation}
  f^k(x,y) = \frac{1}{2\pi \sigma_1 \sigma_2}\exp\left\{-\frac{1}{2}\left[\frac{(x-\mu_1)^2}{\sigma_1^2}+\frac{(y-\mu_2)^2}{\sigma_2^2}\right]\right\},
\end{equation}
where the expected value $\mu_1$ and $\mu_2$ denote the 2D coordinates, and $\sigma_1$ and $\sigma_2$ denote the length and width, which can characterize the influence of the vehicles on the signal to a certain extent. As shown in the upper right conner of Fig.~\ref{fig:vqmix}, the feature map $\mathcal{G}$ are divided into four channels for the MS, cars, vans and buses, respectively. Taking all cars for example, the data of car channel is 
\begin{equation}
  \mathcal{G}^{car} = \sum f^k(x,y), k\in \text{car}.
\end{equation}

Besides, the UAV can observe its own location $\bm{w}^U(t)$. As a result, the observation of the UAV at time slot $t$ is 
\begin{equation}
    o^U(t) = \{\mathcal{G}(t),\bm{w}^U(t)\}.
\end{equation}

\subsection{State}
The signal is affected by all the environmental objects, including buildings, roads, and vehicles. However, it is impossible to describe these objects completely. Since the size and locations of the vehicles can determine whether the direct path is blocked and significantly affect the signal strength, the state can include $\mathcal{G}$. Besides, the state can also include the location of UAV $\bm{w}^U$. We believe that the size and location information of vehicles and the UAV is sufficient for the hypernetwork to output the weights of mixing network.
In summary, the state is 
\begin{equation}
  s(t) = \{\mathcal{G}(t),\bm{w}^U(t)\}.
\end{equation}
\subsection{Action}
\subsubsection{MS}
The action of the MS at time slot $t$ is $a^M(t)$, which indicates whether the MS choose the direct link or the UAV-assisted link. Specifically, $a^M(t)=1$ indicates that the MS choose the UAV-assisted link, and $a^M(t)=0$, otherwise.
\subsubsection{UAV}
The UAV should choose the flying direction $\bm{a}^U(t)$ at each time slot. For simplicity, the flying direction is discretized into five directions: left, right, forward, backward, and hover, i.e., $\{(-1,0),(1,0),(0,1),(0,-1),(0,0)\}$.
\subsection{Reward}
Since the MS and UAV have the same objective \eqref{pro:dis}, they share the common reward. Considering that $R^M(t)$ is a relatively large variable, we define the reward as
\begin{equation}
  r(t)=\kappa(R^M(t)-\bar{R}-\Gamma(t)),
\end{equation}
where $\bar{R}$ can be considered as the mean value of $R^M(t)$, and $\kappa$ is the positive ratio between reward and the throughput. Obviously, maximizing $\sum^T r(t)$ is equal to maximizing \eqref{pro:dis}. Compared with the direct optimization of \eqref{pro:dis}, the reward can focus on the difference among $R^M(t)$.
\subsection{Network Architecture}
As shown in Fig.~\ref{fig:vqmix}, the MS agent first extract the feature of $\mathcal{F}(t)$ through a ResNet \cite{resnet}, and then the extracted feature is flattened and concatenates the MS location $\bm{w}^W(t)$. A GRU cell takes the concatenated feature and the hidden state of the last time slot $h_2^s(t-1)$ as the input and output the new hidden state $h_2^s(t)$. The state-action value $Q_2(\eta_2,\cdot)$ is generated by a fully connected network. Then the MS agent chooses action through $\epsilon$-greedy policy. Specifically, the MS chooses a random action with probability $\epsilon$, and chooses the action with maximum value with probability $1-\epsilon$. The UAV agent has a network architecture similar to the MS agent.

The hypernetworks extract the feature of the state through a ResNet, and generate the weights of the mixing network through multilayer perceptrons (MLPs). The mixing network inputs the state-action value of two agents, and outputs $Q_{tot}$.
\section{Performance Evaluation}
\subsection{Simulation Setting}


We adopt the CARLA \cite{carla}, an autonomous driving simulation platform, to simulate the environment. We focus on the road as shown in Fig.~\ref{fig:system}, which is $85$ meters long, and $15$ meters wide. There are three types of vehicles, i.e., \emph{car type}, \emph{van type}, and \emph{bus type}, whose length, width, and height are [3.71\;m, 1.79\;m, 1.55\;m], [5.20\;m, 2.61\;m, 2.47\;m], [11.08\;m, 3.25\;m, 3.33\;m] respectively \cite{weihua}. The types of vehicles are randomly generated, and we utilize a traffic simulation software, SUMO \cite{sumo}, to simulate the speed and trajectory of all the vehicles. The MS belongs to the car type. The driving time of the MS $T$ is determined by the road length and the trajectory of the MS. The time slot length is $\delta_t=0.05$\;s. The flying altitude of the UAV is $h^U=10$\;m and the antenna height of the BS is $h^B=3$\;m. The number of subcarrier is $K=16$. The transmit powers of the UAV and BS on each sub carrier are $P^U_k=0.1$\;W and $P^B_k=1$\;W, respectively. The total bandwidth is $KB_k=100$\;MHz, and the noise power spectral density is $n_0=10^{-17}$\;W/Hz. The initial location of the UAV is $\bm{w}_0^U=(0,0)$.

The channel dataset is generated by Wireless Insite \cite{wi}, a ray tracing software. The carrier frequency is set as $28$\;GHz, and the propagation model is X3D. The building material is concrete, and the vehicle material is metal. We set the maximum number of reflections as $6$, the maximum number of diffraction as $1$. Wireless Insite can compute the paths between the transmitter and the receiver, and the channel can be computed according to \eqref{eq:channel}. We compute all the channels between the BS and the points along the MS trajectory to get all the $\{H^{BM}_k\}$. Considering the UAV trajectory optimization, we need to compute all channels between the UAV and the MS aa well as between the UAV and the BS at all possible locations. Specifically, we divide the road into $3\times 35 = 105$ grids with grid length $2.5$\;m, since the UAV flying direction is discretized and the flying speed is constant. The UAV can fly one grid per time slot. We compute all the channels between the $105$ grid points and the points along the MS trajectory $H^{UM}_k$ and between the grid points and the BS $H^{BU}_k$.
\subsection{Performance and Analysis}
In order to show the importance of vision, the baseline method only makes decisions based on the location information without visual information.

\begin{figure}
  \centering
  \includegraphics[width=0.4\textwidth]{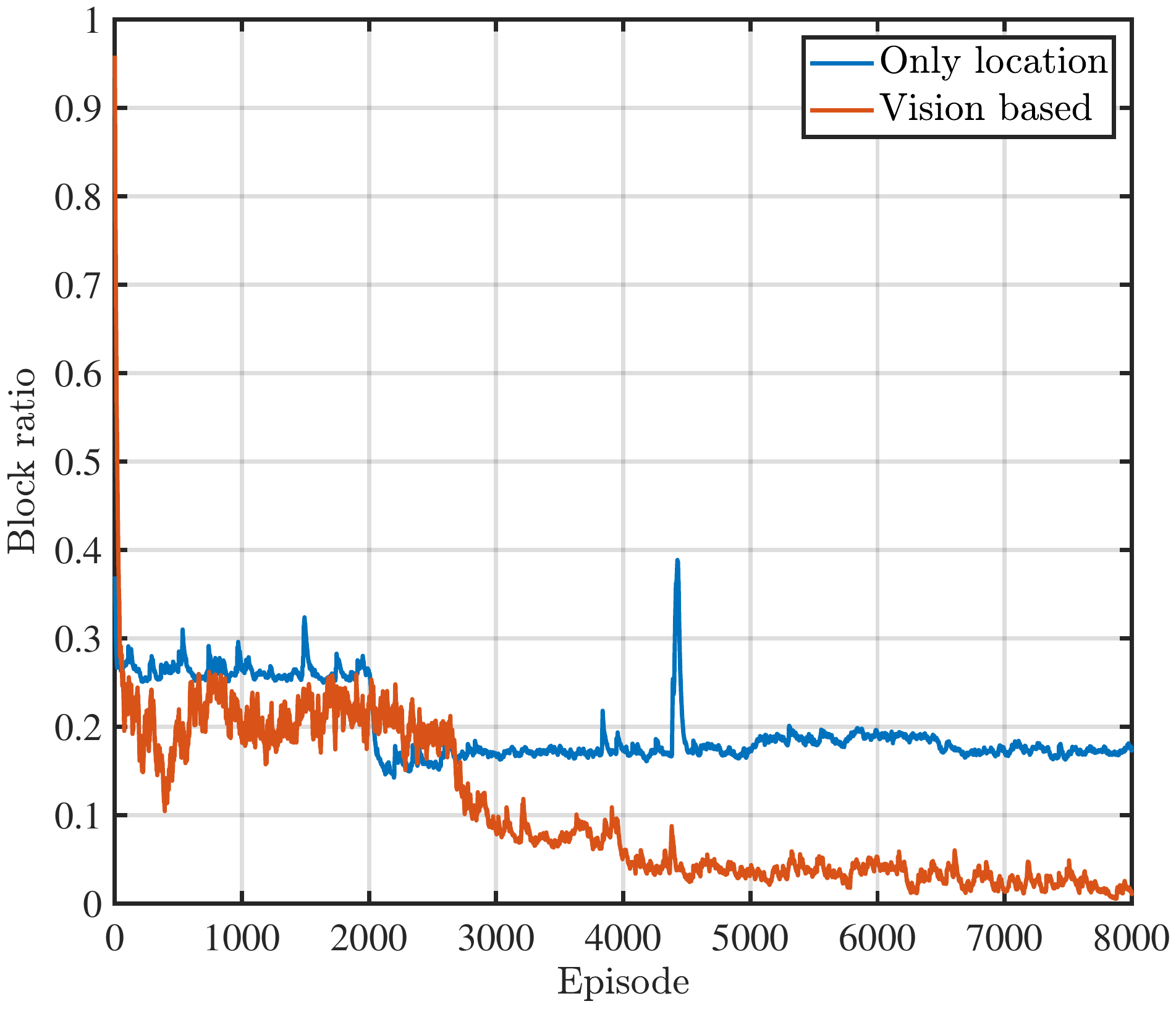}
  \caption{Block ratio.}
  \label{fig:block_ratio}
\end{figure}

First, we show the block ratio performance in Fig.~\ref{fig:block_ratio}. In this figure, we can find that the block ratios of both methods decrease with the training process, and the VQMIX based on visual information outperforms the baseline. This is because the vision includes the location and the size information of the vehicles around the MS, which basically determines whether the link is blocked. With the visual information, the MS can choose the link properly to avoid the blockage. Besides, with only location information, the block ratio may also significantly decrease, because the MS can choose the UAV-assisted link to reduce the block ratio at the cost of delay.

\begin{figure}
  \centering
  \includegraphics[width=0.4\textwidth]{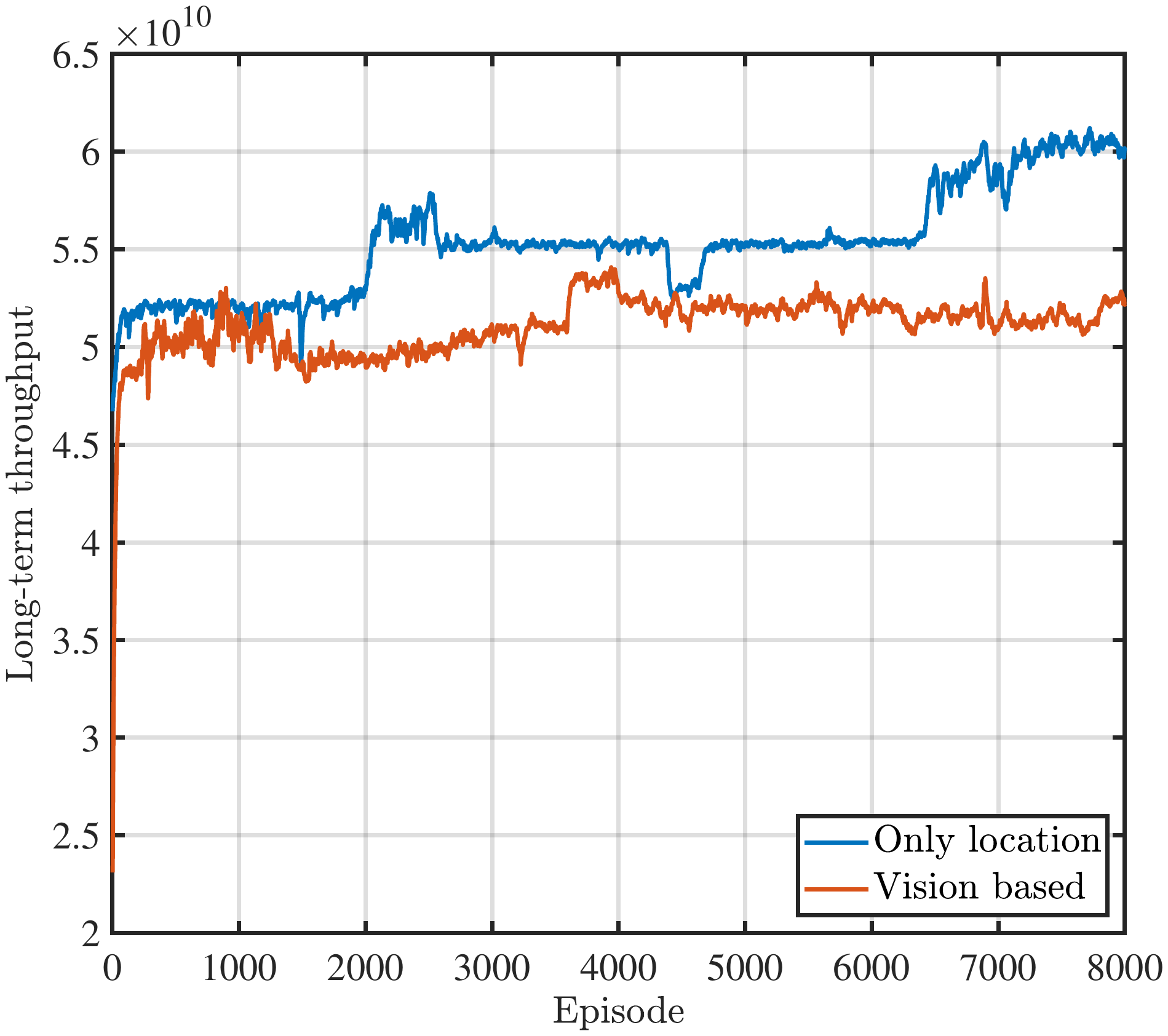}
  \caption{Long-term throughput.}
  \label{fig:throughput}
\end{figure}

Next, we provide the long-term throughput in Fig.~\ref{fig:throughput}. We find that the VQMIX based on visual information can achieve higher long-term throughput compared to the baseline with only location information. This is because VQMIX can achieve lower block ratio to prevent severe attenuation of the received signal power, which results in higher SNR and long-term throughput.

\section{Conclusion}
In this paper, we propose a vision-aided handover framework for UAV-assisted V2X communications, to leverage visual information to make decisions between direct link and UAV-assisted link for enhancing the throughput. Simulation results has demonstrated that the proposed VQMIX can avoid blockage caused by vehicles around the MS with visual information and improve the throughput.

\appendices



\ifCLASSOPTIONcaptionsoff
  \newpage
\fi



%
\bibliographystyle{IEEEtran}
\bibliography{IEEEabrv,ref}
\end{document}